# Niobium and niobium nitride SQUIDs based on anodized nanobridges made with an Atomic Force Microscope

M. Faucher[1,2], T. Fournier[1], B. Pannetier[1], C. Thirion[3], W. Wernsdorfer[3], J.C. Villegier[4] and V. Bouchiat[1]

[1]CRTBT, UPR CNRS 5001, BP 166, 38042 Grenoble, France.
[2]GPEC, UMR CNRS 6631 Université de la Méditerranée, Case 901, 13288 Marseille, France.
[3]Laboratoire Louis Néel, UPR CNRS 5051, BP 166, 38042 Grenoble, France.
[4]DRFMC, CEA-Grenoble, 17 rue des Martyrs, 38054 Grenoble Cedex 9, France.

## ABSTRACT

We present a fabrication method of superconducting quantum interference devices (SQUIDs) based on direct write lithography with an Atomic Force Microscope (AFM). This technique involves maskless local anodization of Nb or NbN ultrathin films using the voltage biased tip of the AFM. The SQUIDs are of weak-link type, for which two geometries have been tested: Dayem and variable thickness nanobridges. The magnetic field dependence of the maximum supercurrent $I_c(\Phi)$ in resulting SQUIDs is thoroughly measured for different weak link geometries and for both tested materials. It is found that the modulation shape and depth of $I_c(\Phi)$ curves are greatly dependent on the weak link size. We analyze the results taking into account the kinetic inductance of nanobridges and using the Likharev-Yakobson model. Finally we show that the present resolution reached by this technique (20nm) enables us to fabricate Nb weak-links which behavior approaches those of ideal Josephson junctions.

**Corresponding author** : Vincent Bouchiat, email : bouchiat@labs.polycnrs-gre.fr
Centre de Recherches sur les très basses températures
CNRS, BP. 166, F-38042 Grenoble Cedex, France
Tel : +33 4 76 88 10 20  Fax : +33 4 76 87 50 60

# Introduction

Lithography using scanning probe microscopes (SPM) [1-2] is an emerging technology, which enables surface patterning with a nanometer scale resolution. Indeed, the typical linewidth is beyond the range of those obtained using conventional lithographies which are based on exposition of a suitable resist with *far field* emitted photons or electrons . Since SPMs operate in the *near field* regime, they show a greatly reduced proximity effect [3], an important drawback that limits resolution in electron beam lithography. Furthermore, SPM lithography techniques have brought during the last decade some new features to device fabrication, such as easy alignment and *in-situ* control of the device electrical characteristics during its fabrication [2].

Among them, those involving an atomic force microscope (AFM) are now the most widely used because they show better versatility and do not involve ultra-high vacuum technology. Actually AFM microscopes can provide different ways for surface patterning, each method involving a different physical phenomena. As an example, one can cite the tip indentation patterning that takes advantage of the local tip-surface mechanical interaction leading to a surface scratching [4-6]). Another technology is given by the local electrochemical oxidation of the surface using a voltage biased tip [9-17]. In this paper, we will focus on this latter technique although SQUIDs have been successfully obtained using the former one [6].

Indeed, local anodization of the surface of a semiconductor [8,9] or of non-noble metals [10-12] by the biased tip of an AFM is now a popular method for fabrication of nanoscale quantum devices. Quantum point contacts on a 2D electron gas [8] or in metal [10], nanowires [9], single electron devices [11,17], superconducting devices [13] as well as other nanoscale devices involving nanotubes [14] or clusters [15] have been obtained with this technique.

In this paper, we present a first application of this technique to the fabrication of weak-link Superconducting Quantum Interference Devices (SQUIDs). We show that the design of superconducting weak links can be controlled in three dimensions, leading to nanometer-sized variable thickness bridges (VTB).

## **Material and techniques**

Local anodization induced by scanning probe microscopes has been pioneered by Dagata et al. in the early 90's [7]. A Silicon surface was first used, but the process was rapidly extended to numerous non-noble metals (Cr, Al, Ti, Nb,…). Its principles are quite simple : the tip of the AFM is scanned over the surface to be patterned. A water bridge is formed at the tip/sample interface and an electrochemically-induced oxidation of the film takes place in this tiny water volume (Fig. 1) when the tip-sample bias exceeds a negative threshold.

Pulsed biasing have shown to provide a better height/width aspect ratio [17] for the oxide lines with respect to DC biasing, while operating the AFM in intermittent contact mode, preserves the metal covered tip from rapid erosion [18] with respect to contact mode for which large electrostatic forces are superimposed to the tip.

This direct-write patterning technique can be used as a single "negative" fabrication step without further processing since the oxidized areas becomes insulating. However, in order to fabricate the nanostructure in a single step, the film thickness must be less than the depth of oxidation, which is typically of the order of 10nm. Therefore, as a first stage, one must obtain superconducting ultra-thin film of Nb and NbN of thickness 4-10 nm having good crystalline quality.

These films are deposited on an annealed sapphire wafer. A pre-patterning using conventional UV photolithography and reactive-ion-etching has been performed. Details of the film fabrications and preliminary results are described elsewhere [19].

Low temperature measurements of their transport properties are summarized in Fig. 2. Electrical transport in these films shows superconducting properties depressed with respect to the bulk [20], mostly for the thinnest films (6nm and below). However the superconductivity is still well established at liquid helium temperature for 6nm-thick films made of both materials. This thickness has been chosen for their compatibility with the lithographic procedure and is kept constant for all the results presented in the following. Typical critical current density are $3.10^7$ A/cm² for 6nm-thick Nb films at 0.05K, while its resistance in the normal state is about 30 Ω/square. For the same films, a residual resistivity ratio ($R_{300K}/R_{4K}$) around 2 indicates that the low temperature mean free path is of the order of 15 nm for the Nb films.

AFM Lithography is directly performed on the strip lines after a careful cleaning procedure in a NaOH (1M) aqueous solution in order to remove surface contaminants that can strongly affect the lithography. A negative voltage ranging between 4 and 14V is applied on a commercial Pt-Ir covered tip with respect to the grounded film. Usual tip speed is about 100nm/s leading to a 30 nm wide oxide line. Large insulating areas are obtained by scanning the biased tip in lines laterally separated by 10 - 40 nm . This helps to define a loop, which acts as an input coil for the SQUIDs. The loop has an inner area *A* of about 1µm² for all devices, and geometric inductance *L* of each arm of the loop is estimated to be about *0.8pH*.

Two type of weak link geometries have been elaborated: lateral constrictions leading to Dayem bridges [21] (typical dimensions 50 nm wide, 300nm long) similar to those studied in [22] and variable thickness bridges (VTB), obtained by drawing a thin oxide line across Dayem bridges which partially oxidizes the thickness of the niobium layer (see Figs. 5b and 6). VTB SQUIDs are of particular importance because their "three-dimensional" configuration lead to a better geometrically defined and better thermalized weak link with respect to Dayem SQUIDs. Such a nanometer scale 3D engineering of nanobridges remains an important feature specific to this technology.

## Low temperature SQUIDs measurements

All tested SQUIDs show at low temperature hysteretic *I-V* characteristics [19], see Fig.3. As the current is ramped from zero, the SQUID transits to a finite voltage branch at a switching current $I_c$, which is the main characteristics of the device. Transition is caused by the propagation of a hot spot from the weak-links. Hysteresis is attributed to a slower cooling during the ramp down which bring back the SQUID in the superconducting state for currents much lower than $I_c$. A specific detection technique has been elaborated in order to optimize the precision and speed of $I_c$ measurements [22]. The modulations of $I_c$ with an externally applied magnetic field *B* have been measured for different temperatures in the range [0.04-5K]. For all tested devices a quasi-periodic modulation of $I_c$ is observed with a periodicity of around 20 Gauss (see Fig. 4), which is in good agreement with the predicted period $\Phi_0/A$, where $\Phi_0$ is the superconducting flux quantum. The modulation is washed out at high fields (several hundreds of Gauss) and the critical current is strongly attenuated. The width of the attenuation envelope directly depends on the weak link geometry. As presented in Fig. 4, it is directly correlated to the weak link surface *S* (see inset Fig. 3) exposed to the magnetic field *B*. While the modulation periodicity remains constant due to a loop area maintained at $A=1\mu m^2$, the curve envelope is affected the well-known diffraction effect, that usually occurs when a flux quantum penetrates a Josephson junction of finite area [23].

However, unlike for an ideal SQUIDs, the $I_c(B)$ modulations present a triangular shape dependence and the modulation depth show a strongly reduced modulation depth ($\frac{\Delta I_c}{I_c} \approx 15\%$), much lower than expected. Furthermore, some SQUIDs show multivaluated switching currents. All these features are usually encountered in damped SQUIDs with large geometric inductances for which efficient screening currents wash out the interference pattern [23] and trap into the loop different integer numbers of flux quanta. In our case however, the geometric inductance cannot account for this effect since the screening factors $\beta=\dfrac{2\pi\, 2LI_c}{\Phi_0}$ are all in the range [0.03-0.1], which is negligible compared to 1.

On the other hand, our devices are dominated by the relatively large kinetic inductance $L_K$ of the weak links [24,25] associated to the the kinetic energy of the superconducting condensate in the weak links. It has a dominating contribution in a weak link which size is large compared to the superconducting coherence length [22,26,27]. The model developed in the following shows that within certain limits, a large kinetic or a large geometric inductance can have the same influence on the SQUID $I_c(B)$ curves.

## **Interpretation and discussion**

The phase relation within the loop can be written as [23]:

$$2\pi\left(n-\dfrac{\Phi_{ext}}{\Phi_0}\right)=\varphi_1-\varphi_2+\dfrac{2\pi\, LI_1}{\Phi_0}-\dfrac{2\pi\, LI_2}{\Phi_0} \qquad (1),$$

for which $\Phi_{ext}=B\cdot A$ is the external applied magnetic flux, $n$ is an the integer number of flux quanta in the loop, $\varphi_1$ and $\varphi_2$ are the phase differences across each junction. $I_1$ (resp. $I_2$) are the currents in each arm of the SQUID.

On the other hand, we have:

$$I=I_1+I_2 \quad (2)$$

One can define the $F$ function by $F(I)=\varphi(I)+\dfrac{2\pi\, LI}{\Phi_0}$ .

Equation (1) becomes $2\pi(n-f)=F(I_1)-F(I_2)$ \hspace{2cm} (1').

The maximum value of I given by the set of equations (1') and (2) leads to the experimentally measured flux modulation $I_c(B)$. Let us find the expression of $F$ for the two cases of Josephson junctions in series with a large geometric inductance $L$ (case 1) and for a weak link without inductance but with a length $l$ much longer than the superconducting coherence length $\xi$ (case 2).

*Case 1* : For an ideal Josephson junction, the current phase-relation is sinusoidal : $I = I_c \sin\varphi$

$F$ becomes : $F(I) = \arcsin(\frac{I}{I_c}) + \beta \frac{I}{I_c}$ where $\beta$ is the already introduced screening factor.

For large inductances, the screening factor satisfies the relation $\beta \gg 1$, therefore the second term dominates over the first one, and $F$ becomes a linear function of $I$.

*Case 2* : the current-phase relation for a long ($l \gg \xi$) weak link is derived from the non-linear Ginzburg-Landau equation. Following the Likharev-Yakobson model [25], it gives for the current-phase relation :

$$I = \frac{\Phi_0}{2\pi L_K}\left(\varphi - \frac{\xi^2}{l^2}\varphi^3\right) \qquad (3),$$

where $L_K$ is the kinetic inductance of the bridge. It is related to the length $l$ and cross section $\sigma$ of the bridge by the expression [23] :

$$L_K = \mu_0 \lambda^2 \frac{l}{\sigma} \qquad (4).$$

Therefore, if one introduces by analogy with $\beta$, the kinetic inductance screening factor $\beta_K = \frac{2\pi L_K I_c}{\Phi_0}$, one has :

$$I = \frac{I_c}{\beta_K}\left(\varphi - \frac{\xi^2}{l^2}\varphi^3\right) \qquad (5).$$

For $\beta_K > 1$ the solution of eq. 1 becomes multivaluated [25], and the phase $\varphi$ spans only a limited range before switching to the neighboring state. Therefore the current $I$ becomes a linear function of $\varphi$. The function $F$ have then the following expression: $F(I) = \beta_K \frac{I}{I_c}$, which is a complete analogy with case 1 by exchanging $\beta$ and $\beta_K$. This justifies why a weak link SQUID with a large kinetic inductance has a similar flux dependence as an damped SQUID. For both type of devices, the total maximum current shows a linear dependence in $\Phi_{ext}/\Phi_0$ [mod ½], with a slope $dI_c/d\phi$ which equals to $\pm 1/\beta$ and $\pm 1/\beta_K$ respectively. This slope is refered in the following as the SQUID sensitivity .

One can check that our Dayem SQUIDs fulfill the condition $\beta_K >1$, as seen in Fig. 4. Typical kinetic inductances $L_K$ in Nb are around 100-700 pH (see Fig. 8), which is more than 2 orders of magnitude more than the geometric loop inductance $2L$. As predicted by Meservey [24], $L_K$ is expected to scale with the geometric factor $l/\sigma$, where $l$ is the weak link length and $\sigma$ its cross section. In order to verify the complete dependence of our SQUIDs with $L_K$, we have measured the invert of the SQUID sensitivity for VTB Nb SQUIDs of different geometric factors (see Fig. 8). A linear behavior as expected from formula (4) is indeed observed if one takes for the bridge length $l$ the total length of the weak link.

The temperature dependence of $I_c$ and of its flux derivative $dI_c/d\phi$ (the so-called SQUID sensitivity ) have been measured for typical Nb and NbN Dayem bridges (see Fig. 7). The reduced SQUID sensitivity of NbN SQUIDs is attributed to the very small superconducting coherence length of that material (3nm) which leads, for a given geometry, to kinetic inductance $L_K$ larger than in the Nb case thus to a smaller $dI_c/d\phi \propto 1/\beta_K$. The predicted T dependence of $dI_c/d\phi$ (T) can be obtained from eq. 4 assuming for $\lambda$ a temperature dependence given by the "two-fluid" model [28] : $\lambda \propto 1/\sqrt{1-(T/Tc)^{1/4}}$ . Therefore one has :

$$dI_c/d\phi(T) \propto \lambda^{-2} \propto 1-(T/T_c)^{-1/4}$$

The measured dependence is in good agreement with this prediction taking for the critical temperature of the Nb and NbN films 4.1 and 4.5K respectively.

Finally, we have measured SQUIDs with VTBs fabricated using the best resolution presently reached by our technique. One founds that a noticeable and reproducible deviation from a linear regime is observed when one shrinks the weak link size down to the 20 nm range (figure 5, bottom).

This is caused by the dimensions of the weak links which becomes comparable with the superconducting coherence length $\xi$. Indeed, a measurement of the dependence of the critical temperature $T_c$ with the magnetic field have lead to $\xi \sim 10$nm in these films. When one shrinks the wink-link size, a continuous distorsion from the linear regime towards the regime of SQUIDs with ideal Josephson junction $I_c \sim |\cos \Phi|$ appears [25].

Furthermore the modulation depth of small VTB SQUID is almost doubled with respect to Dayem SQUIDs. Thus the kinetic inductance has been reduced to values about 1. It is another experimental signature of smaller junctions.

Applications of these nanobridge SQUIDs can be numerous, as larger Dayem SQUIDs obtained using electron beam lithography [29] have already been used in many different fields : mesoscopic physics with the measurement of persistent currents in 2D electron gas rings [30]. SQUID microscopy with high resolution vortex imaging [31], and nanomagnetism with the measurement of the magnetization reversal in nanoscale ferromagnetic particles [32].

Improvements in the technology such as the recent development of nanotube-terminated AFM tips that brings the resolution of local anodization well below 10nm [33] could help to design tunnel barriers showing a genuine Josephson effect.

**Acknowledgments**

We are indebted to V. Safarov, A. Benoit, and K. Hasselbach for help and discussions. One of us (M.F) thanks the D.G.A. for grant support.

# REFERENCES

[1] for a review on early developments in STM lithography, see R. Wiesendanger, Appl. Surf. Sci. 54 (1992) 271.

[2] E. S. Snow and P. M. Campbel, Science, 270 (1995) 1639.

[3] K. Wilder, B. Singh, D. F. Kyser, and C.F. Quate, J. Vac. Sci. Technol. B, 16 (1998) 3864, see also M. Ishibashi, S. Heike; H. Kajiyama, Y. Wada and T. Hashizume, Jpn. J. Appl. Phys, 37 (1998)1565.

[4] L. L. Sohn and R. L. Willet, Appl. Phys. Lett. 67(1995)1552.

[5] V. Bouchiat and D. Estève, Appl. Phys. Lett. 69(1996)3098.

[6] B. Irmer, R.H. Blick, F. Simmel, W. Gödel , H. Lorenz, and J.P. Kotthaus, Appl. Phys. Lett. 73 (1998) 2051

[7] J.A. Dagata, J. Schneir, H.H. Harary, C.J. Evans, M.T. Postek, J. Bennett, Appl. Phys. Lett. 56, (1990) 2001.

[8] R. Held, T. Heinzel, P. Studerus, K. Ensslin, M. Holland, Appl. Phys. Lett. 71(1997) 2689, R. Held, T. Vancura, T. Heinzel, K. Ensslin, M. Holland, W. Wegscheider, Appl. Phys. Lett. 73, (1998) 262.

[9] E.S. Snow and P.M. Campbell, Appl. Phys. Lett. 64(1994)1932.

[10] E.S. Snow, D.Park and P.M. Campbell, Appl. Phys. Lett. 69(1996)269.

[11] K. Matsumoto, M. Ishii, K. Segawa, Y. Oka, B. J. Vartanian, and J. S. Harris, Appl. Phys. Lett. 68(1996)34, see also K. Matsumoto . Appl. Phys. Lett. 72 (1998) 1893.

[12] B.Irmer, M. Kehrle, H.Lorentz, and J.P. Kotthaus, Appl. Phys. Lett. 71(1997)1733.

[13] I. Song, B.M. Kim, and G. Park, Appl. Phys. Lett. 76 (2000) 601.

[14] P. Avouris, T. Hertel, R. Martel, T. Schmidt; H.R. Shea, and R.E. Walkup, Appl. Surf. Sci., 141(1999)201.

[15] R.J.M. Vullers, M. Ahlskog, M.Cannaerts, C. van-Haesendonck, Appl. Phys. Lett.76 (2000) 1947.

[16] K. Matsumoto, Y. Gotoh, J. Shirakashi, T. Maeda, J. S. Harris, Proceedings of IEDM 97, IEEE p155, (1997).

[17] K. Matsumoto, Y. Gotoh, T. Maeda, J. Dagata, J. Harris, Appl. Phys. Lett. 76 (2000) 239.

[18] M . Tello, R. Garcia, Appl. Phys. Lett. 79 (2001) 424.

[19] V. Bouchiat, M. Faucher, T. Fournier, B. Pannetier, C. Thirion, W. Wernsdorfer, Appl. Phys. Lett. 78 (2001) 160, for NbN films see: J.C. Villegier, N. Hadacek, S. Monso, B. Delaet, A. Roussy, P. Febvre, G. Lamura, J.Y. Laval, IEEE Trans. Appl. Super., 11 (2001) 68

[20] S.I. Park and T.H. Geballe, Physica B 135 (1985) 108.

[21] P.W Anderson and A.H. Dayem, Phys. Rev. Lett. 13 (1964) 195 , A.H. Dayem, Appl. Phys. Lett. 9 (1966) 47.

[22] K. Hasselbach, D. Mailly, J.R. Kirtley, condmat /0110517, submitted to J. Appl. Phys.

[23] A Barone & G. Paterno, "Physics and applications of the Josephson effect", ed. Wiley, (1982).

[24] Meservey, R. and Tedrow, P. M., J. Appl. Phys. 40 (1969) 2028.

[25] K.K Likharev, Rev. Mod. Phys. 51(1979)101.

[26] W. Richter and G. Albrecht, Phys. Status Solidi A,17 (1973) 531.

[27] C.M. Falco, and W.H. Parker, J. Appl. Phys. 46 (1975) 3238.

[28] M. Tinkham, "introduction to superconductivity", ed. Mc Graw-Hill, NY (1996).

[29] C. Chapelier, M. El-Khatib, P. Perrier, A. Benoit, D. Mailly, Proceedings of the 4[th] International Conference SQUID'91, Springer Verlag, Berlin, (1991).

[30] C. Chapelier et al. Phys. Rev. Lett. 70 (1993) 2020.

[31] K. Hasselbach et al. Physica C 332 (2000)140, C. Veauvy et al. condmat /0110517.

[32] W.Wernsdorfer, E.B. Orozco, K. Hasselbach, A. Benoit; B. Barbara, N. Demoncy, A. Loiseau, H. Pascard, D. Mailly, Phys. Rev. Lett. 78 (1997)1791.

[33] Y.Gotoh, K. Matsumoto, T.Maeda, E.B. Cooper, S.R. Manalis, H. Fang, S.C. Minne, T. Hunt, H. Dai, J. Harris, C.F.Quate, JVST A 18, (2000)1321.

# Figure Captions

Figure 1: Principles of the anodization of metallic ultrathin films using an Atomic Force microscope. While scanning a negatively biased tip onto the film, a nanometer size oxide line is directly grown onto the film. Thickness, linewidth and composition of the patterned oxide depend on tip bias voltage and speed.

Figure 2 top: Superconducting onset critical temperature of Nb (circles) and NbN (squares) thin films as a function of the film thickness.
bottom : Residual resistivity ratio (R=300K/R=4K) for the same Nb films.

Figure 3: *I-V* characteristics of a typical Nb Dayem SQUIDs fabricated with the AFM measured at 40 mK using current bias . The switching current Ic is defined as the maximum observed Josephson current. Inset : AFM micrograph of a typical Nb Dayem SQUID of weak link surface *S* and loop area *A*. Patterned areas on the film appear in bright on the topography since they are Niobium oxide protrusions. Notice the atomic steps on both patterned and non-patterned areas, signature of a high crystalline quality.

Figure 4: Magnetic field modulation of the normalized switching current $I_c^*$ (defined by the switching current divided by its maximum recorded value) as a function of the applied magnetic field for Dayem SQUIDs with different weak-link surfaces *S*. The temperature is in the range 40-50mK. Curves have been vertically shifted for clarity.
From top to bottom, *S* equals $8.10^{-3}$, 0.06, 0.15 and 0.4 µm², respectively. The kinetic inductance $\beta_K$ screening factors deduced from the modulation depth are respectively : 3.7, 8.4, 7.8, and 15.

Figure 5: Dependence of the switching current with the applied magnetic field ($T=40$mK) for SQUIDs based on each geometry. (a): For SQUIDs with 0.3µm-long "Dayem"-like bridges, a perfect saw-tooth modulation is obtained, since bridges have width and length larger than the superconducting coherence length $\xi$.
(b): For Nb SQUIDs with VTB of length 20nm, the magnetic field dependence of Ic shows a deviation from the linear behavior (dotted lines) while modulation depth is increased with respect to curve (a). These features appear as signatures of Josephson junctions of dimensions shorter than $\xi$.

Figure 6: AFM micrograph of a VTB SQUID obtained on a 5nm-thick Niobium nitride film. It involves a 3D AFM patterning. In figure a), the two nanobriges are obtained by laterally oxidizing the two sides of the loop following a "notch" pattern. b) Micrograph of the same SQUID for which the oxidized parts have been removed by dipping the sample in an alkali solution. c) Zoom of the right VTB weak-link showing the short transversal groove and the notch pattern.

Figure 7: Temperature dependence for typical Nb (circles) and NbN (squares) Dayem SQUIDs of the sensitivity ($dI_c/d\phi$) (top) and of the maximum critical current (bottom). For comparision the critical currents are divided by the maximum value $I_0$ measured at the lowest temperature. The temperature dependence of the sensitivity is fitted using the kinetic inductance formalism and the predicted dependence of $\lambda$ (see text).

Figure 8: Dependence of kinetic inductance $L_K$ of Nb VTB SQUIDs with the bridge geometric factor $l/\sigma$ at 50mK.

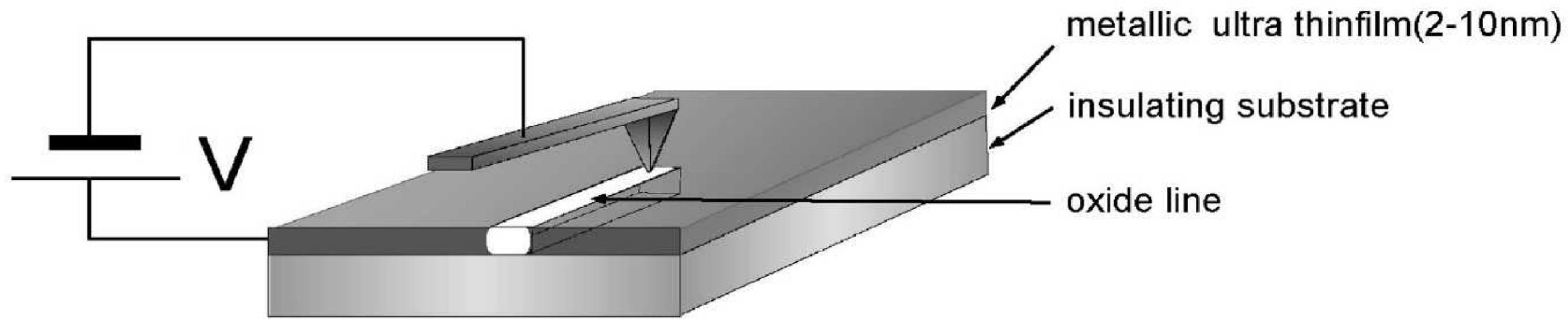

Figure 1

FAUCHER et al.

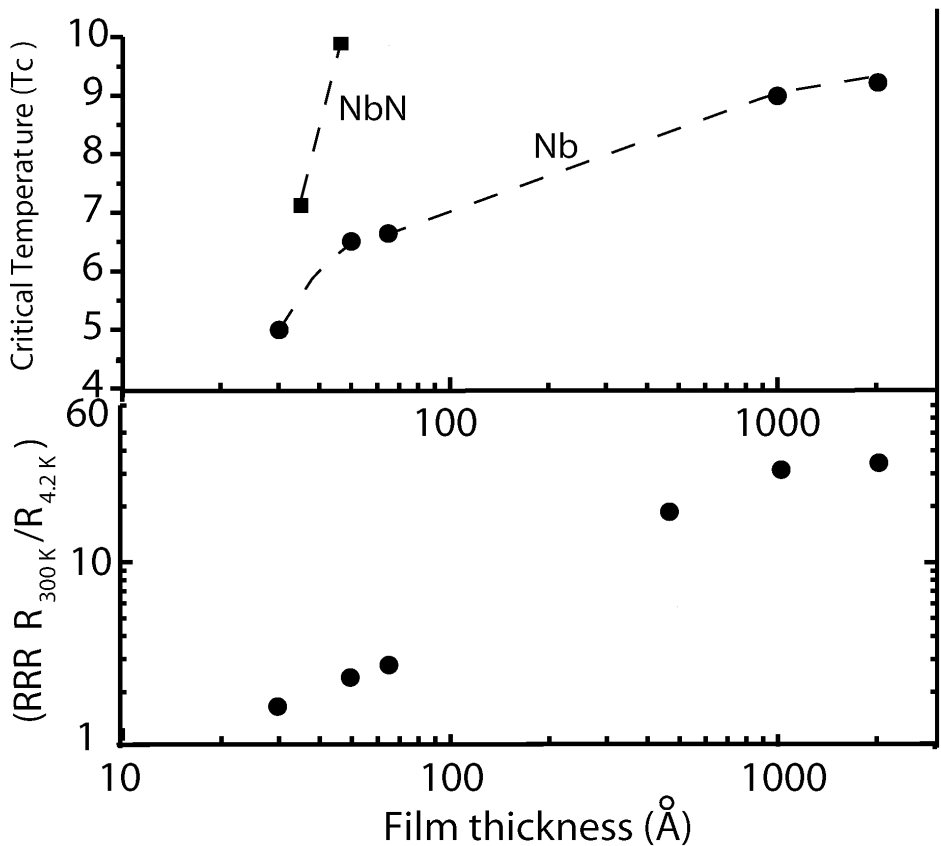

Figure 2
Faucher et al.

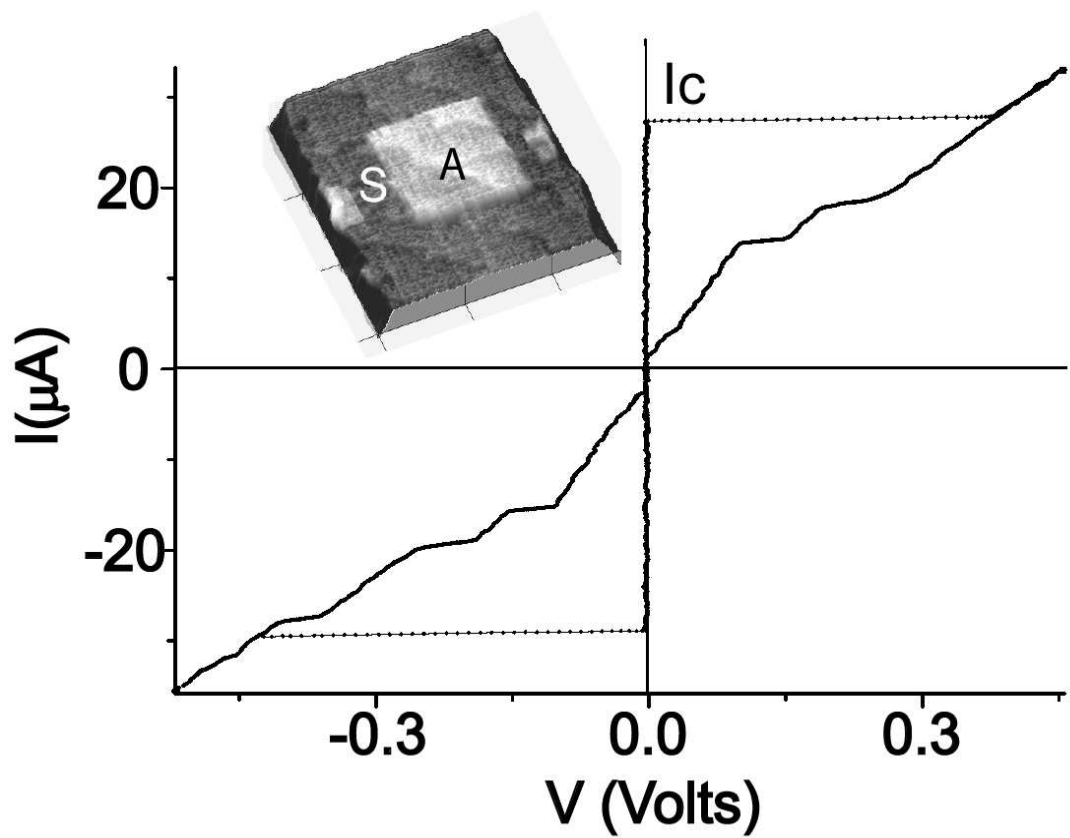

Figure 3
Faucher et al.

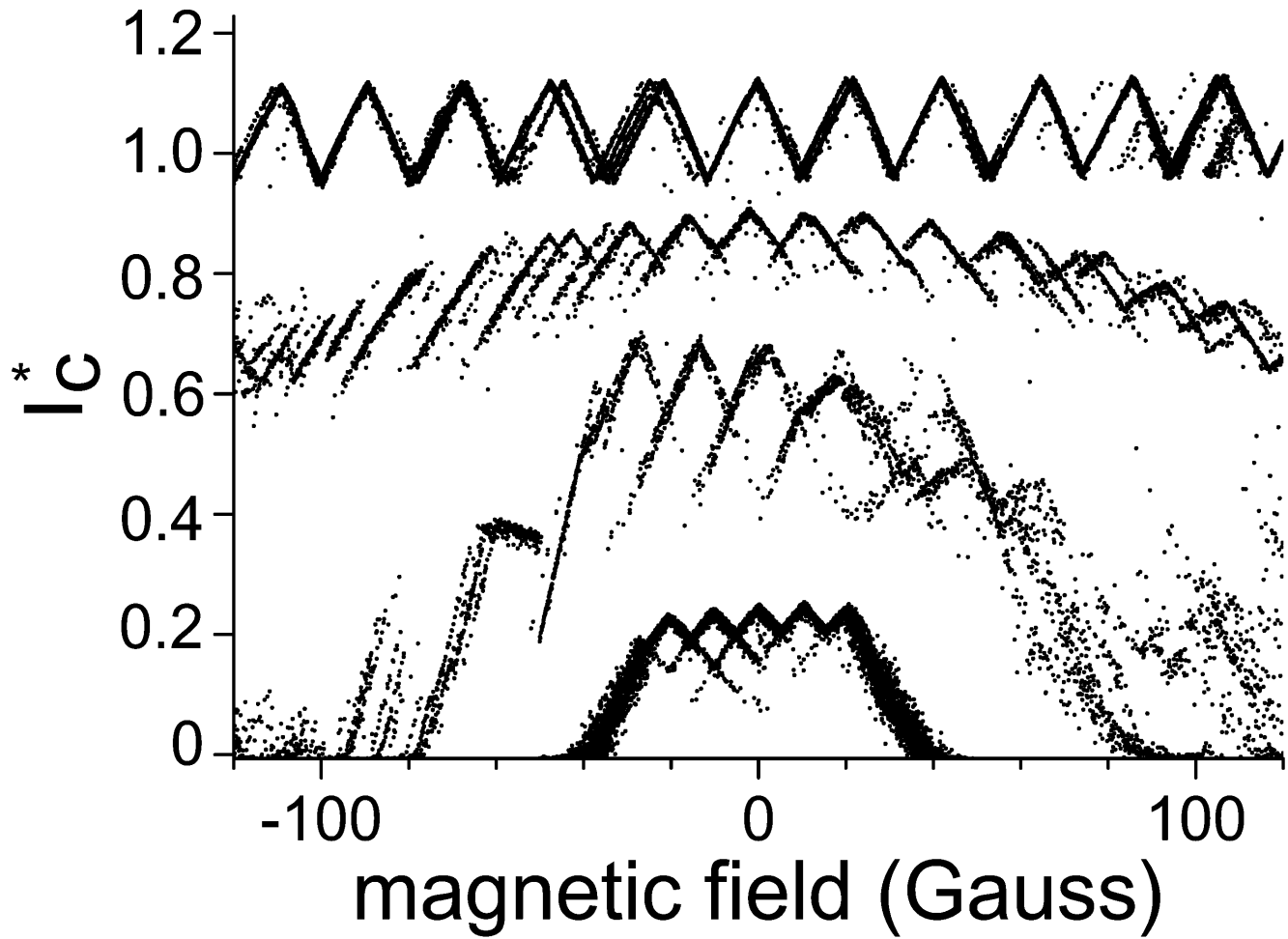

Figure 4 , Faucher et al.

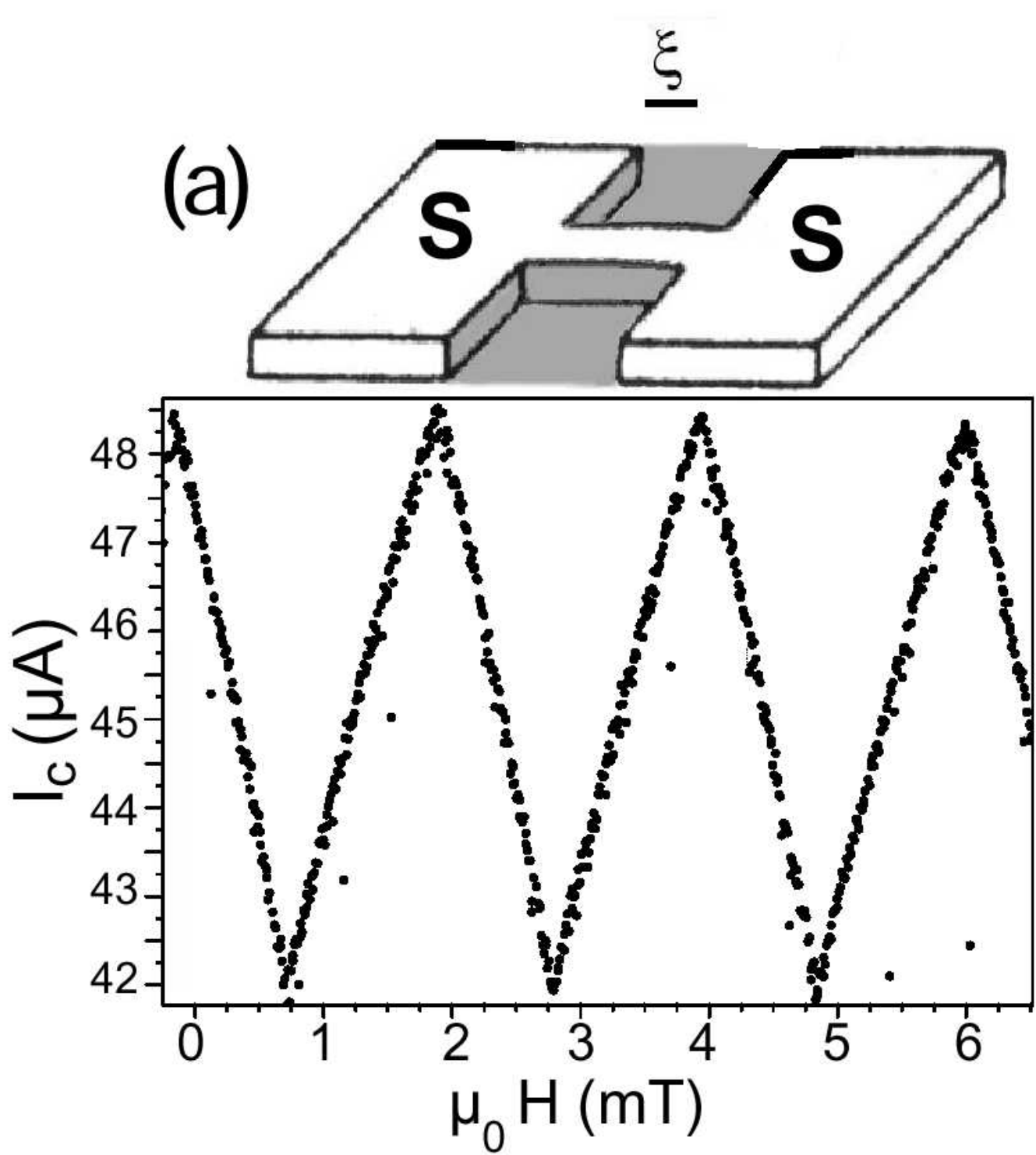

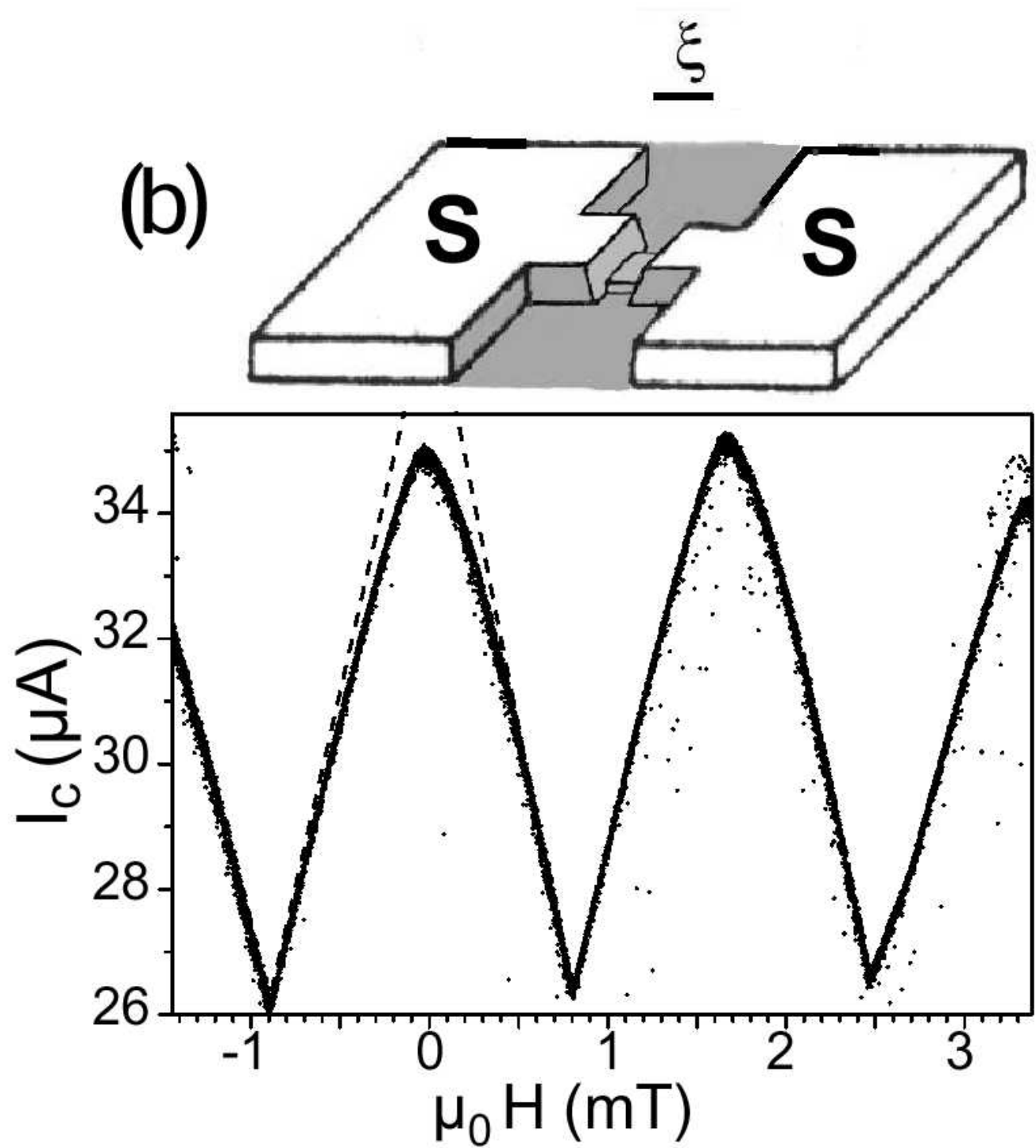

FIG. 5            Faucher et al.

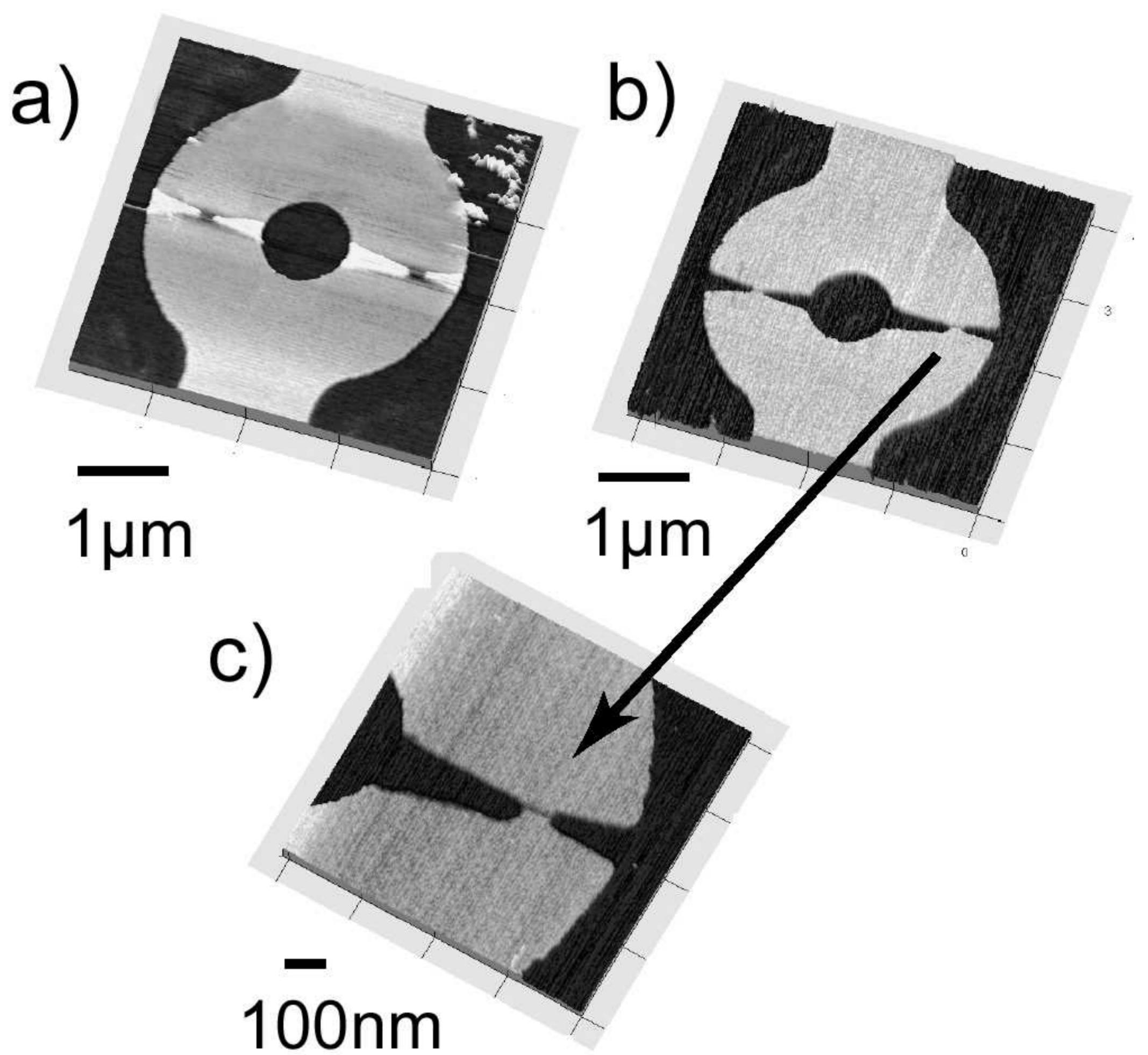

Figure 6

Faucher et al.

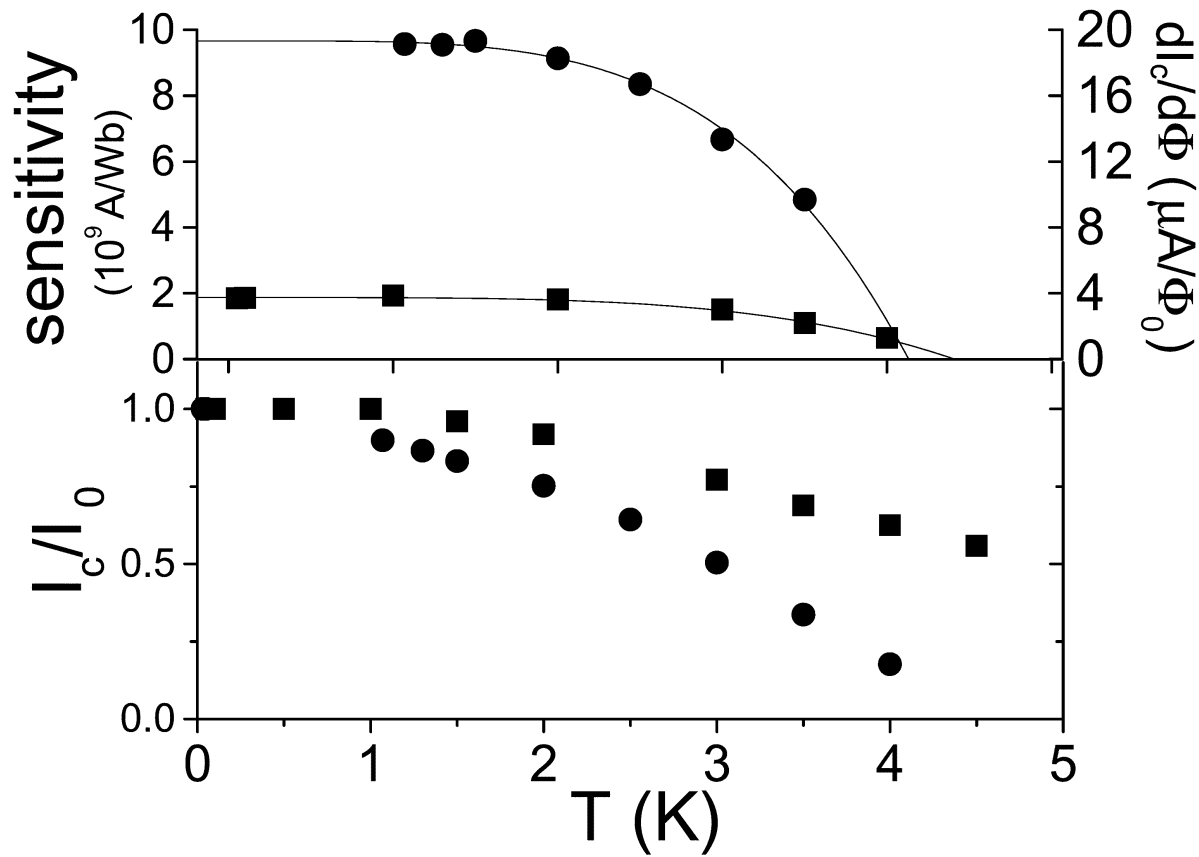

Figure 7
Faucher et al.

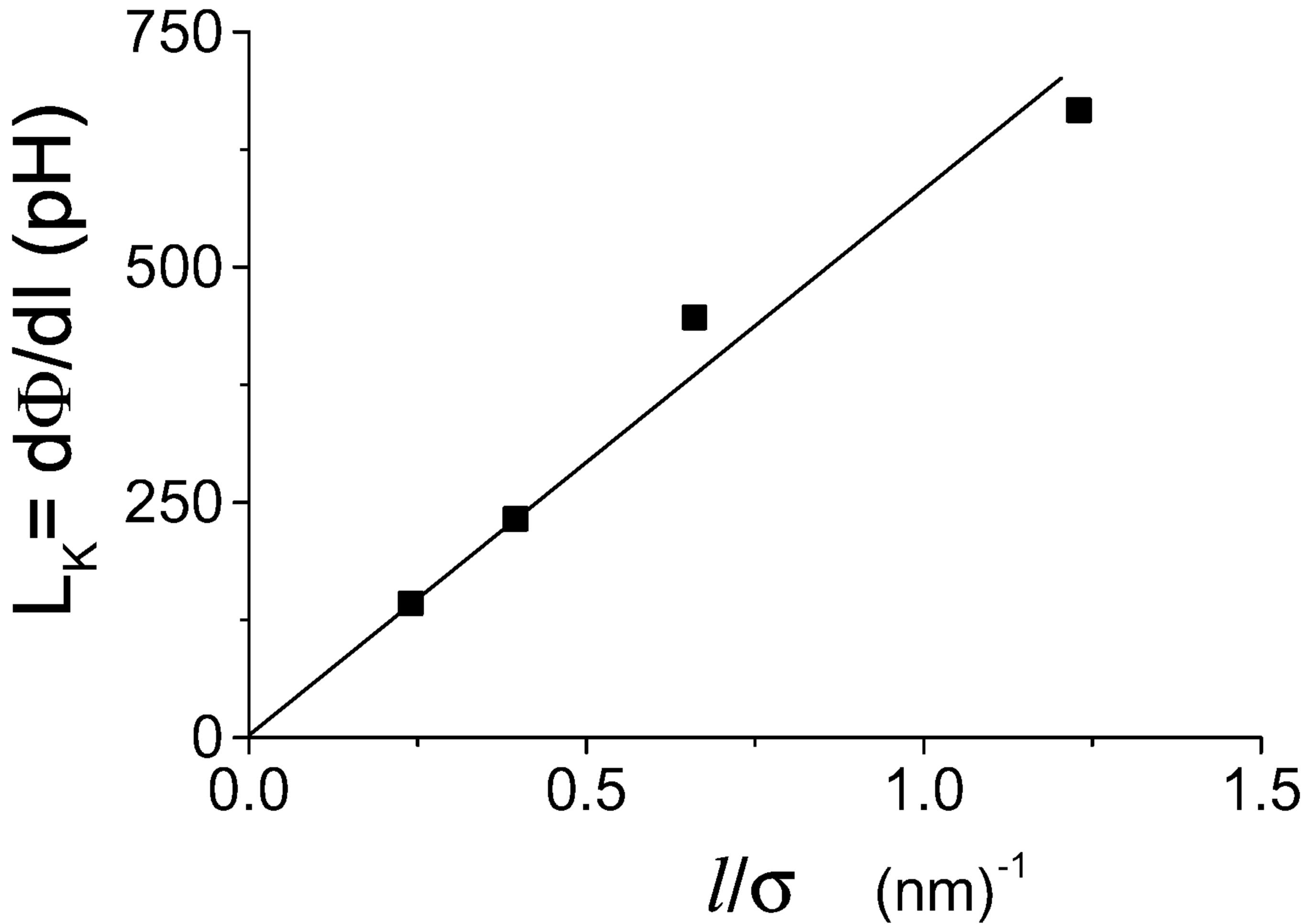